\documentclass[aps,twocolumn,showpacs,amsmath,amssymb,superscriptaddress]{revtex4-1}

\usepackage{graphicx}
\usepackage{color}
\usepackage{subfigure}
\usepackage{dcolumn}
\usepackage{bm}
\usepackage{epsf}
\usepackage{amsmath,amstext,amssymb}
\usepackage{enumitem}
\usepackage{hyperref}
\usepackage[ansinew]{inputenc}
\usepackage{braket}
\PassOptionsToPackage{numbers,sort&compress}{natbib}

\makeatletter
\g@addto@macro\normalsize{%
  \setlength\abovedisplayskip{3pt}
  \setlength\belowdisplayskip{3pt}
  \setlength\abovedisplayshortskip{3pt}
  \setlength\belowdisplayshortskip{3pt}
}
\makeatother

\newcommand{\be}{\begin{equation}}
\newcommand{\ee}{\end{equation}}
\newcommand{\bea}{\begin{eqnarray}}
\newcommand{\eea}{\end{eqnarray}}

\begin{document}

\title{Phase-stable free-space optical lattices for trapped ions}


\author{C.~T.~Schmiegelow}
\author{H.~Kaufmann}
\author{T.~Ruster}
\author{J.~Schulz}
\author{V.~Kaushal}
\author{M.~Hettrich}
\author{F.~Schmidt-Kaler}
\author{U.~G.~Poschinger}\email{poschin@uni-mainz.de}

\affiliation{Institut f\"ur Physik, Universit\"at Mainz, Staudingerweg 7, 55128 Mainz, Germany}

\begin{abstract}
We demonstrate control of the absolute phase of an optical lattice with respect to a single trapped ion. The lattice is generated by off-resonant free-space laser beams, we actively stabilize its phase by measuring its ac-Stark shift on a trapped ion. The ion is localized within the standing wave to better than 2\% of its period. The locked lattice allows us to apply displacement operations via resonant optical forces with a controlled direction in phase space. Moreover, we observe the lattice-induced phase evolution of spin superposition states in order to analyze the relevant decoherence mechanisms. Finally, we employ lattice-induced phase shifts for inferring the variation of the ion position over 157~$\mu$m range along the trap axis at accuracies of better than 6~nm.

\end{abstract}


\maketitle

\noindent
Confinement of laser-cooled neutral atoms by means of the ac-Stark effect from standing waves (SW) generated by laser interference has been established two decades ago \cite{WEIDEMUELLER1995,BIRKL1995}. This has enabled the realization of quantum-gas microscopes \cite{BAKR2009,SHERSON2010} and the development of ultra-accurate atomic clocks \cite{NICHOLSON2015}. For trapped ions, laser interference patterns can be employed for exerting spin-dependent optical dipole forces, which are most prominently used for entangling gate operations \cite{SORENSEN2000,LEIBFRIED2002}. 

Phase stable optical lattices offer interesting applications for trapped ion systems, ranging from micromotion-free confinement to tunable periodic potentials for quantum simulations. Furthermore, deep lattices can serve for the investigation of friction models \cite{Pruttivarasin2011,BYLINSKII2015}. In recent experiments, localization of trapped ions in SWs of optical cavities has been realized \cite{MUNDT2002,LINNET2012,LINNET2014,KARPA2013}, as well as trapping of ions in optical fields \cite{ENDERLEIN2012}. While cavity-based systems serve to create deep lattices with inherent phase stability, free-space laser fields as a complementary approach allow for rapid switching and modulation of the SW. Recent proposals for experiments in the field of quantum thermodynamics \cite{DORNER2013,GOOLD2014} and non-equilibrium phase transitions \cite{GENWAY2014} can be realized by harnessing free-space phase-stable optical lattices. 

\begin{figure}[h!tp]\begin{center}
\includegraphics[width=0.42\textwidth]{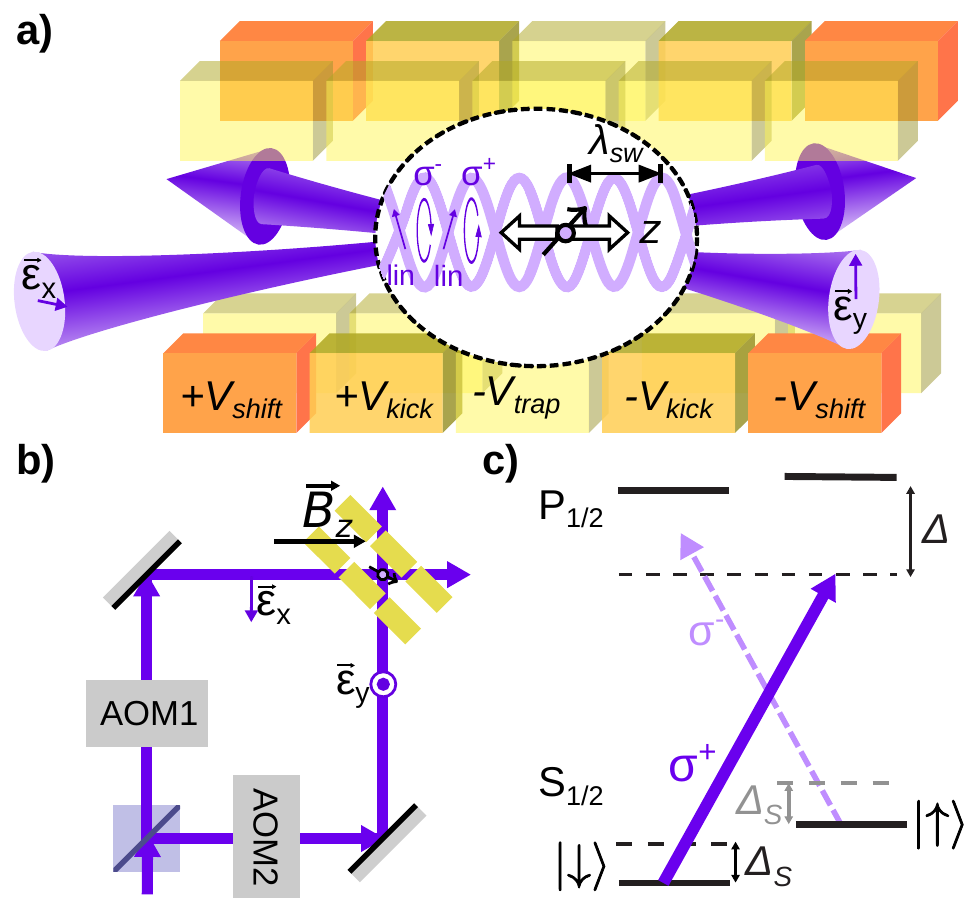}
\caption{ (color online) Realization of phase-stable standing waves for trapped ions. \textbf{a)} An ion trapped in a segmented Paul trap is exposed to two laser beams with polarizations $\varepsilon_{x,y}$ generating an interference pattern of alternating lin$\perp$lin polarization. The ion is trapped by applying a voltage on the center segment (yellow), kicked by sudden voltage changes in the neighbouring segments (dark-yellow), shifted along the trap axis by changing the voltage at the next-nearest neighbour trap segments (orange). \textbf{b)} Sketch of the interferometer comprised of the two laser beams, controlled by acousto-optical modulators, and the trapped ion. \textbf{c)} Relevant energy levels of $^{40}$Ca$^+$. The lattice beams off-resonantly address the $S_{1/2}\leftrightarrow P_{1/2}$ transition, inducing a position dependent ac-Stark shift of the ground state sub-levels, alternately shifting the  $\ket{\downarrow}$ and $\ket{\uparrow}$ as the ion is displaced along the standing wave.
}
 \label{fig:sketch}
\end{center}\end{figure}

In this work we employ off-resonant laser beams to generate a free-space optical lattice which we actively stabilize by measuring its induced ac-Stark shift on a ground state cooled ion. The lattice beams comprise a Mach-Zehnder interferometer, where the ion is to be considered as its closing element. We use the phase-stable lattice to demonstrate \textit{phase-controlled} spin-dependent displacement operations mediated by optical forces. 

Moreover, we observe and model the relevant spin-decoherence mechanisms for a trapped ion exposed to the SW, by characterizing the phase evolution of a spin superposition state at different positions. We find that phase jitters are the dominant decoherence mechanism at nodes of the SW, while decoherence at the antinodes is mainly caused by thermal fluctuations.  Finally, we demonstrate how measurements of the lattice-induced ac-Stark shifts over a wide position range of about 100$\mu$m along the trap axis serve for tracking the ion position with an accuracy far below the optical wavelength, and even better than the extension of the quantum mechanical wavepacket.

The experimental setting is depicted in Fig. \ref{fig:sketch}. We trap a single $^{40}$Ca$^+$ ion in a segmented, linear Paul trap \cite{SCHULZ2008}. The Zeeman sublevels of the ground state $\ket{\downarrow}\equiv\ket{S_{1/2},m_J=-\tfrac{1}{2}}$ and $\ket{\uparrow}\equiv\ket{S_{1/2},m_J=+\tfrac{1}{2}}$ define a two-level system\cite{POSCHINGER2009}, which is split by $2\pi\cdot$13.26~MHz by an external magnetic field aligned at 45$^{\circ}$ to the trap axis. Population in $\ket{\downarrow}$ is detected with an accuracy of better than 99\% by observing fluorescence on the S-P transition after electron shelving the population in $\ket{\downarrow}$ to the metastable $D_{5/2}$ state. Coherent rotations between $\ket{\downarrow}$ and $\ket{\uparrow}$, side-band cooling of axial motion near the ground state and determination of the motional excitation are carried out by driving stimulated Raman transitions mediated by two off-resonant laser beams near 397~nm \cite{POSCHINGER2009,POSCHINGER2010}.

The micro-structured trap allows us to control the ion's axial position by biasing voltage segments nearby the trapping segment, see Fig.\ref{fig:sketch}a. We use this both to displace the ion by a sudden voltage change\cite{WALTHER2012} and for precise positioning of the ion in the SW. The electrode voltages are controlled by a field programmable gate array based arbitrary waveform generator\cite{WALTHER2012,RUSTER2014} with a voltage resolution of $0.3$~mV, a minimum update rate of $400$~ns and a timing resolution of $20$~ns. By monitoring the ion position $x$ versus shift voltage $V_s$ on an EMCCD camera, we determine $dx/dV_s\vert_{x=0}=$8~$\mu$m/V, which yields a positioning resolution of 2.4nm.

The SW is generated by two laser beams near $\lambda=\,$~397~nm, detuned by $\Delta=2\pi\cdot$30~GHz from the $S_{1/2}\leftrightarrow P_{1/2}$ transition. Both beams are derived from the same laser source and are switched and modulated by single-pass acousto-optical modulators as indicated in Fig. \ref{fig:sketch}. The power in each beam is roughly 2~mW, and the spot size at the trap center is about 210~$\mu$m. The beams are aligned as indicated in  Fig. \ref{fig:sketch}, resulting in a SW aligned along the trap axis with a lin$\perp$lin configuration \cite{COURTOIS1992}. Hence, the polarization is periodically varying from left circular to right circular (at the antinodes) via linear (at the nodes) with a period of $\lambda_{sw}=\lambda /(2\sin(\alpha/2))$, where $\alpha$ is the angle enclosed by the two beams. Circularly polarized optical fields give rise to a differential energy shift between $\ket{\uparrow}$ and $\ket{\downarrow}$. We thus obtain a \textit{position dependent} ac-Stark shift:
\begin{equation}
\Delta_S(z)=\Delta_S^{(0)} \cos(kz+\phi),
\label{eq:deltaS}
\end{equation}
where $k=2\pi/\lambda_{sw}$, and $\Delta_S^{(0)}=\Omega_1\Omega_2/4\Delta$ is the ac-Stark shift amplitude determined by the dipolar Rabi frequencies $\Omega_{1,2}$ pertaining to each beam. 

The relative phase of both optical fields $\phi$ is subject to interferometer drifts. We stabilize $\phi$ by measuring the ac-Stark shift via a spin-echo experiment and feeding back on the ion position. Exposed to the SW for a time $t$, a superposition state of an ion located at $z$ acquires a phase shift $\ket{\uparrow}+\ket{\downarrow}\rightarrow \ket{\uparrow}+e^{i\Delta_S(z) t}\ket{\downarrow}$. By repeating this measurement 200 times, we obtain an estimate of the probability 
\begin{equation}
\mathcal{S}(z,t)=\tfrac{1}{2}\left[1+\cos(\Delta_S(z) t)\right]
\label{eq:spinechosignal}
\end{equation}
to detect the ion to be in $\ket{\uparrow}$. We stabilize the ion to a position half way between a node and an antinode. For maximum sensitivity, we choose $t$ such that a $\pi/2$ phase shift is acquired at this point, yielding a setpoint signal $\mathcal{S}_{set}=1/2$. From this, in conjunction with the feedthrough $dx/dV_s$, we obtain an error signal to feedback on the ion position. Monitoring this signal in the locked state over 10~min, we observe residual phase fluctuations of $\Delta\phi=$0.03$\pi$. The experiments presented in the following are carried out by interleaving the phase stabilization sequence with a \emph{science} experiment, which allows for updating the lock every $\approx$~0.5s. During the actual experiment sequence, the ion can be reliably placed at any position in the SW by changing the shift voltage. The lock performance is close to the limit imposed by readout shot-noise which we estimate to be $\Delta\phi=2\sqrt{\mathcal{S}_{set}\left(1-\mathcal{S}_{set}\right)/N}\approx$0.02$\pi$ at the setpoint. We observe and correct for total phase drifts of about $3\pi$ within 10~min and maximum drift rates of about 0.03$\pi$/s, which confirms both the necessity and the feasibility of the active stabilization scheme. 

\begin{figure}[htp]\begin{center}
\includegraphics[width=0.48\textwidth]{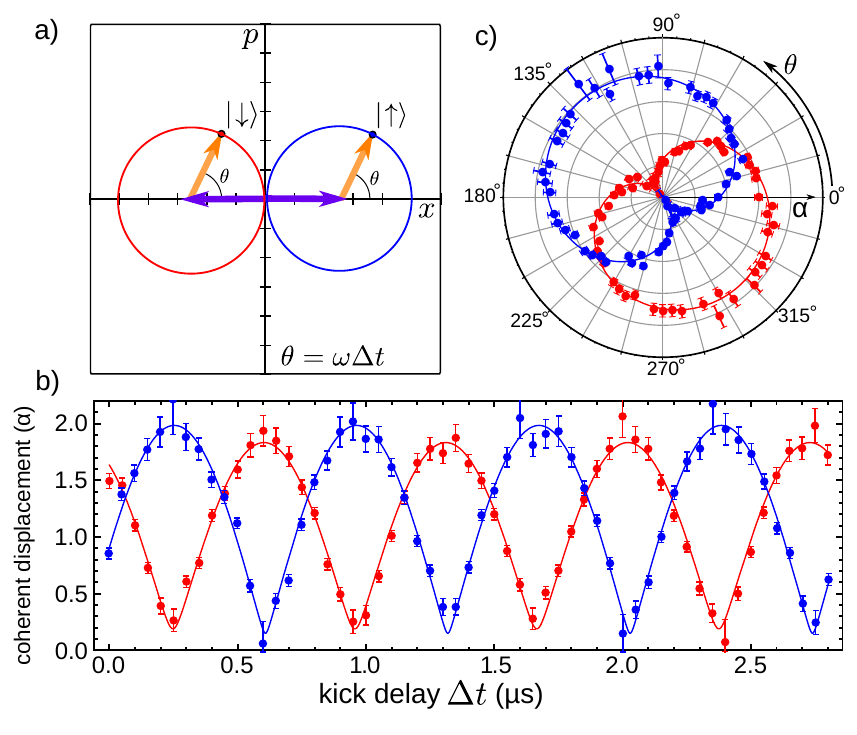}
\caption{ (color online) Phase coherent combination of spin-dependent optical and electrical displacement operations. \textbf{a)} Phase-space picture of the optical kicks (purple) and electrical kicks (orange). The direction of the optical kick is given by the spin, while the direction of the electrical kick is determined by the delay time. \textbf{b)} Measured motional excitation vs. delay of the electrical kick for both spin initializations (blue: $\ket{\uparrow}$, red:$\ket{\downarrow}$). Solid curves are fits to Eq. \ref{eq:finaldisplacment}. The signals for opposite spin initializations are out of phase by $\pi(1.02\pm0.04)$. The displacement amplitude is inferred from the result  probing both red and blue sideband at a pulse area of $\pi$, each at 200 repetitions. We take into account an initial thermal excitation corresponding to $\bar{n}\approx$0.4, as measured independently.  \textbf{c)} Displacement versus the electrical kick phase in polar phase space coordinates.}
 \label{fig:phasestablekicks}
\end{center}\end{figure}




The phase stabilization scheme allows for displacing the ion in phase space, with control over the oscillation phase. By detuning the SW beams with respect to each other by the trap frequency $\omega\approx2\pi\cdot$1.41~MHz via the modulators (see Fig. \ref{fig:sketch}b), we generate a slowly  \textit{running wave} (RW) which gives rise to a spin-dependent resonant optical force. This RW inherits the phase stability of the locked static SW. Exposing the ion to the optical force for a time $t$ causes an oscillator displacement of amplitude $\alpha(t)=-i\,m_J\eta\Delta_S^{(0)}t e^{i(kz+\phi)}$, where $\eta\approx$0.21 is the Lamb-Dicke factor. The phase-space direction of the optical force depends on spin direction $m_J$ and on the position of the ion in the RW. Therefore, by controlling the absolute phase of the RW, we control the \textit{direction} of the optical kick. We demonstrate such control by displacing the ion back to rest with an auxiliary electrical field kick.

The experimental sequence starts with a sideband cooled ion initialized in $\ket{\uparrow}$ by optical pumping, or in $\ket{\downarrow}$ by using a $\pi$ pulse thereafter. Then, an optical kick is applied, where the amplitude and duration are chosen to yield a displacement amplitude of $\vert\alpha_0\vert\approx$1. After a variable delay time $\Delta t$, we exert an electrical kick by biasing the neighboring segments of the trap segment transiently to $\pm V_k$, where $V_k$ and the pulse time are adjusted to provide the same displacement amplitude $\vert\alpha_0\vert$ as the optical kick. The phase $\theta$ between optical kick and the electric kick is controlled by the wait time: $\theta=\omega \Delta t$. As illustrated in Fig. \ref{fig:phasestablekicks}a, the expected displacement magnitude is given by
\begin{equation}
\vert\alpha(t)\vert^2=2\vert\alpha_0\vert \left(1+2 m_J\cos(\theta+\phi)\right).
\label{eq:finaldisplacment}
\end{equation}
To probe the resulting displacement, we drive either the red or blue motional sideband of the stimulated Raman transition \cite{WALTHER2012}. 

The results are shown in Fig. \ref{fig:phasestablekicks} show that the electrical and optical kicks can reliably cancel each other within the measurement precision. Flipping the spin changes the phase of the oscillation by $\Delta\theta=1.02(4)\pi$. The electrical and optical kicks can reliably cancel each other: For $\ket{\uparrow}$($\ket{\downarrow}$) we obtain $\tfrac{\alpha_{\text{min}}}{\alpha_{\text{max}}}=0.05(1)\left(0.14(8)\right)$. Thus, phase-stabilized spin-dependent optical forces can be applied to displace the harmonic motion of trapped ions in any desired direction in phase space. 

\begin{figure}[ht!p]\begin{center}
\includegraphics[width=0.49\textwidth]{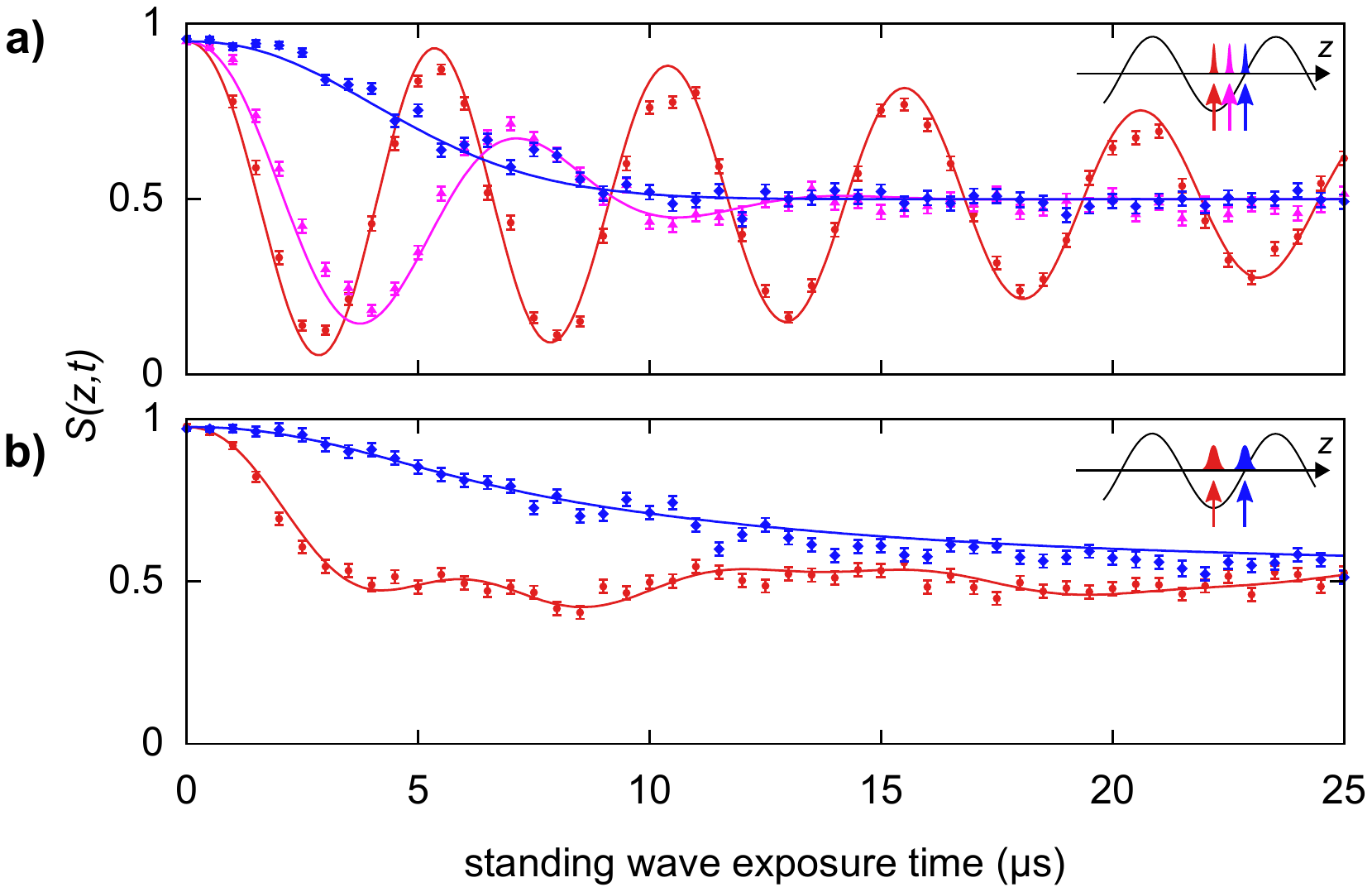}
\caption{ (color online) Temporal phase evolution in the phase-stabilized standing wave. \textbf{a)} Spin-echo signal $\mathcal{S}(z,t)$ versus exposure time $t$ for an ion cooled close to the motional ground state. We probe at an antinode (red), a node (blue) and halfway between (magenta). The solid lines are fits to Eq. \ref{eq:thermalsignal}. We infer a mean phonon number of $\bar{n}=0.4(2)$ and rms phase fluctuations of $\Delta\phi=0.048(3)\pi$. \textbf{b)} Data without groundstate cooling. For the fit we use $\Delta\phi$ obtained from the sideband cooled data and infer a mean phonon number $\bar{n}=28(3)$.}
 \label{fig:timescans}
\end{center}\end{figure}

We now study in detail the spin-decoherence of an ion exposed to the actively stabilized SW by determining the accumulated phase for varying exposure time $t$ by using a spin echo sequence. Measurements are carried out for different fixed positions and with and without employing sideband cooling. 

\begin{figure*}[htp]\begin{center}
\includegraphics[width=0.95\textwidth]{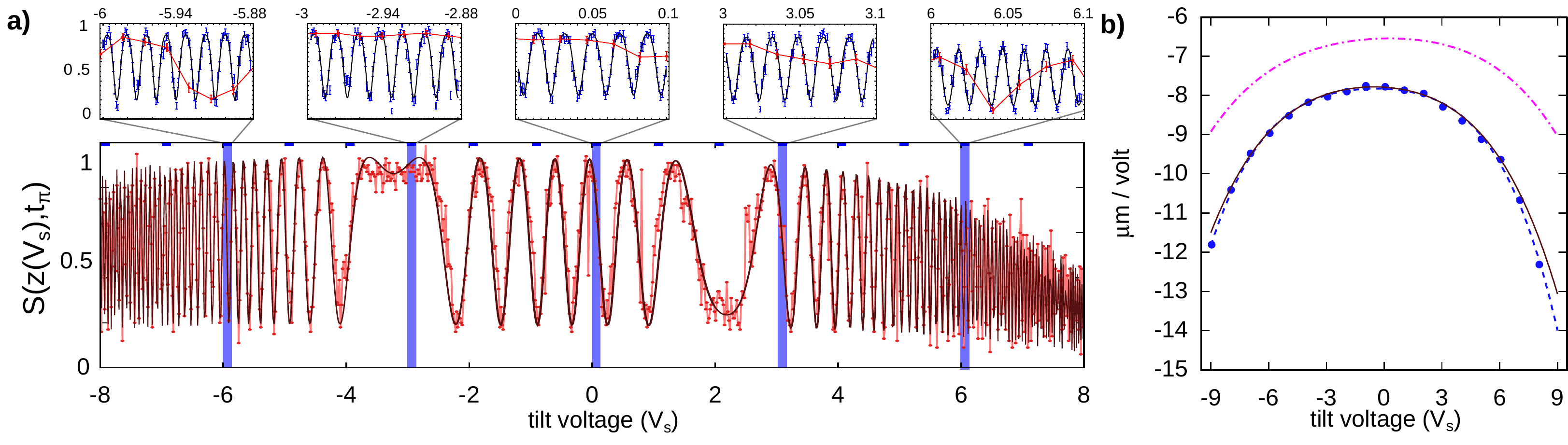}
\caption{ (color online) \textbf{a)}  
Small panels: Spin echo signal vs. shift voltage for resolved scans (blue) in steps of 1.2~mV resolving the SW oscillations. Large panel: Same signal (red) for a step of 18~mV, leading to position steps larger than the SW period, which gives rise to aliasing. The aliased data is also shown in the small panels (red).  All fits shown in black pertain to  Eq. \ref{eq:signal} in conjunction with 5th order polynomial $z_{eq}(V_s)$. The variation of the period is due to the inhomogeneity of the electric field generated by the shift and trap electrodes. \textbf{b)} Variation of $z_{eq}$ with respect to $V_s$, i.e. $z_{eq}'(V_s)$, as inferred from the resolved scans (blue dots), the fit of $z(V_s)$ to these (blue solid) and the fit to the aliased data (black). The magenta curve corresponds to the equilibrium positions as obtained from electrostatic simulations of the trap structure.}
 \label{fig:widescan}
\end{center}\end{figure*}

In the case of a sideband-cooled ion positioned at an antinode, rotations about spin $z$-axis can be seen, where the dephasing is dominated by a residual population of excited harmonic oscillator states, see Fig. \ref{fig:timescans}a. We model this effect by considering that the ac-Stark shift depends on the vibrational quantum number $n$ as $M_n(\eta)\Delta_S(z)$, with the matrix element $M_n(\eta)=\braket{n|\cos(kz)|n}$ \cite{LEIBFRIED2003}. The resulting signal is obtained by averaging the spin-echo signal Eq. \ref{eq:spinechosignal} over a thermal distribution with mean vibrational quantum number $\bar{n}$:
\begin{equation}
\bar{\mathcal{S}}(z,t)=\sum_n \frac{\bar{n}^n}{(\bar{n}+1)^{n+1}} \mathcal{S}_n(t).
\label{eq:thermalsignal}
\end{equation}
This thermally induced decoherence is analogous to the case of spin rotations driven by propagating waves, where the resonant Rabi frequency depends on $n$. Thermal dephasing correctly describes the behavior when the ion is at the node, (Fig. \ref{fig:timescans}a, red). By contrast, we observe a faster decoherence when the ion is located at the antinode (Fig. \ref{fig:timescans}a, blue).  Here,  dephasing from the residual phase jitters of the SW dominates. We model this by averaging the spin echo signal over Gaussian fluctuations of the phase with spread $\Delta\phi$. This yields a \emph{position dependent} contrast-loss function
\begin{eqnarray}
\gamma_n(z,t)&\approx& \left(M_n(\eta) \Delta_S^{(0)} t\right)^2 \tfrac{\Delta\phi^2}{2} e^{-\Delta\phi^2} \nonumber \\
&\cdot& \left(1-\cos(2kz)+\tfrac{3\Delta\phi^2}{2}\cos^2(kz)\right),
\end{eqnarray}
which holds for $\Delta\phi\ll\pi$. The spin-echo signal pertaining to the vibrational quantum number $n$ then reads:
\begin{equation}
\mathcal{S}_n(z,t)=\tfrac{1}{2}\left[1+e^{-\gamma_n(z,t)}\cos(M_n(\eta)\Delta_S(z) t)\right]. 
\label{eq:signal}
\end{equation}
To fit the experimental data shown in Fig. \ref{fig:timescans}, we average $S_n$ according to Eq. \ref{eq:thermalsignal}. The fits reveal phase fluctuations of $\Delta\phi=$0.048(3)$\pi$, slightly worse as compared to the stability characterization. This is due to the fact that only half of the experimental duty cycle is devoted to the stabilization. The phase fluctuations are equivalent to a position fluctuation of $\delta z\approx$0.025$\lambda_{sw}\approx$6.5~nm,  which is smaller than the size of the ground state wave function of the harmonic oscillator of $\sim$18~nm. 

Data for the non-sideband-cooled case is shown in Fig. \ref{fig:timescans}b. Note that the behavior inverts - dephasing is faster at the antinodes than at the nodes, where thermal averaging leads to mitigation of phase fluctuations. At the antinode, the oscillation contrast is lost, which shows that groundstate cooling is required for the stabilization scheme.

We consider a further decoherence mechanism, caused by the \textit{static} spin dependent forces exerted by the SW. At the node, the SW gives rise to the displacement $\alpha=2m_J k\Delta_S^{(0)}\left[\hbar/(2m \omega^3)\right]^{1/2}$. For our typical operation conditions, we calculate $\vert\alpha\vert\approx$0.03, leading to contrast loss of about $e^{-2\vert\alpha\vert^2}\approx$0.2\%. A 30-fold laser intensity would be required to obtain full spin-dependent force induced dephasing.

Finally, the phase-stabilized SW is employed as a precise ruler for position detection. This position detection scheme circumvents the limitations imposed by the resolution of imaging optics or position fluctuations of fluorescing ions \cite{BROWNNUTT2012}. We map out the ion equilibrium position along the trap axis by measuring the spin-echo signal over a wide position range of about 157~$\mu$m. The exposure time $t_{\pi}$ is chosen to correspond to a $\pi$ phase shift at an antinode of the SW. We perform scans where the shift voltage is changed in steps 1.2~mV along windows of size 100~mV, such that the SW is fully resolved, and one aliased scan over the entire voltage range with a scan step of 18~mV. Fig. \ref{fig:widescan}a shows the resulting signals. 

We model the shift of the equilibrium position $z$ with $V_s$ with a polynomial of 5th order, such that the resulting signal is described by $\mathcal{S}(z(V_s),t)$. A nonlinear fit yielding the polynomial coefficients gives relative accuracies for the coefficients $c_i$ on the 10$^{-3}$ level and better, such that the ion position is tracked at an accuracy of better than 6~nm over the whole position range. We compare the results to calculations using the electrostatic potentials $\Phi_i(z)$ pertaining the +1~V applied to electrode $i$, obtained from simulation of the trap \cite{SINGER2010}. For trap voltage $V_t$ applied to segment $n$ shift voltage $\pm V_s$ applied to segments $n\pm 2$, we solve $V_s(\Phi'_{n+2}(z_{eq})-\Phi'_{n-2}(z_{eq}))+V_t\Phi'_n(z_{eq})=0$ for the equilibrium position $z_{eq}(V_s)$. We compare $z'_{eq}(V_s)/\lambda_{sw}$ to the measurement results in Fig. \ref{fig:widescan} b), which reveals discrepancies on the order of 16\% to the simulations. We assume that these originate from imperfections of the trap geometry and electric stray fields.

In conclusion, we have demonstrated active stabilization of a Mach-Zehnder interferometer by means of spin measurements on a trapped ion. We are able to lock the ion position in the SW, which enables accurate position detection over a range determined by the beam focus. This position detection scheme outperforms measurements based on camera detection, both in terms of position resolution and range. The relevant decoherence effects have been quantitatively characterized. The residual phase fluctuations could be further suppressed by using a fiber-coupled interferometer \cite{BALLANCE2014}. We have harnessed the phase stable standing wave to demonstrate phase-referenced displacement operations. The latter might serve as a new tool for quantum state tomography, for reconstruction of quantum states with broken rotational symmetry in phase space, e.g. squeezed states \cite{KIENZLER2015,LO2015}. Phase-stable spin-dependent optical forces might be employed for realizing qubit-state dependent shuttling operations, where trapped ions are moved to different trap sites depending on the qubit state by employing ion separation techniques \cite{RUSTER2014}, while coherence is retained. This might turn out to be an interesting enhancement of the toolbox for scalable quantum logic with trapped ions. Here, positioning a trapped-ion qubit register in a SW could also serve to mitigate single qubit addressing errors: With the parameters from Ref. \cite{SCHINDLER2013} p.25, the crosstalk error for an addressed $\pi$-qubit rotation of about 15\% could be pushed down to the 10$^{-4}$ level with a SW of 2~MHz Stark shift amplitude. Furthermore, the phase-stable lattice can be used to generate quantized ac-Stark shifts, i.e. energy shifts depending on the motional quantum number \cite{SCHMIDTKALER2004}, however without the limitations arising from off-resonant spin rotations. This may serve as a tool for interferometric measurement of quantum dynamics \cite{DORNER2013}. \\
During the work on this manuscript, we became aware of recent related work \cite{DELAUBENFELS2015}. We acknowledge funding by the EU Seventh Framework Program (FP7/2007-2013) under Grant Agreement No. 600645 (SIQS), and by the Bundesministerium f{\"u}r Bildung und Forschung via IKT 2020 (Q.com). UGP acknowledges funding by the Johannes-Gutenberg Universit\"at Mainz. CTS acknowledges support from the Alexander von Humboldt Foundation.

\bibliography{lit}

\end{document}